\documentclass[12pt]{spieman}  
\usepackage{amsmath,amsfonts,amssymb}
\usepackage{graphicx}
\usepackage{setspace}
\usepackage{tocloft} 
\usepackage{lineno}
\usepackage{tikz}
\usepackage{ragged2e}
\usepackage{graphics,graphicx}
\usepackage{natbib}
\usepackage{multicol}
\usepackage{xspace}
\usepackage{amssymb}
\usepackage{url}
\usepackage{wrapfig} 
\usepackage{amsfonts, amsmath, amssymb} 
\newcommand{\chandra}{\textit{Chandra}}

\newcommand{\swift}{\textit{Swift}}
\newcommand{\xmm}{\textit{ XMM-Newton}}

\newcommand{\athena}{{\it Athena}}
\newcommand{\Athena}{{\it Athena}}
\newcommand{\znew}{{}z_{100}}

\newcommand{\apjs}{{\it Astrophys. J. Suppl.}}
\newcommand{\aap}{{\it Astron. \& Astrophys.}}

\definecolor{lime}{HTML}{A6CE39}
\DeclareRobustCommand{\orcidicon}{
	\begin{tikzpicture}
	\draw[lime, fill=lime] (0,0) 
	circle [radius=0.16] 
	node[white] {{\fontfamily{qag}\selectfont \tiny ID}};
	\draw[white, fill=white] (-0.0625,0.095) 
	circle [radius=0.007];
	\end{tikzpicture}
	\hspace{-2mm}
}
\foreach \x in {A, ..., Z}{\expandafter\xdef\csname orcid\x\endcsname{\noexpand\href{https://orcid.org/\csname orcidauthor\x\endcsname}
			{\noexpand\orcidicon}}
}

\title{Exploring rapid transient detection with the \athena\ Wide Field Imager}

\author[a,b,*]{\orcidA{}Pragati Pradhan}
\author[b]{Abraham D. Falcone}
\author[b]{\orcidC{}Jamie A. Kennea}
\author[b]{\orcidD{}David N. Burrows}
\affil[a]{Massachusetts Institute of Technology, 77 Massachusetts Ave. Cambridge, MA 02139, USA}
\affil[b]{Penn State University, Department of Astronomy and Astrophysics, 525 Davey Lab, City, University Park, PA 16802, USA}

\cftpagenumbersoff{figure}
\cftpagenumbersoff{table} 
\begin{document} 
\maketitle
\begin{abstract}
X-ray transients are among the most enigmatic objects in the cosmic sky. 
 The unpredictability and underlying nature of their transient behavior has prompted much study in recent years. While significant progress has been made in this field, a more complete understanding of such events is often hampered by the delay in the rapid follow-up of 
 any transient event. An efficient way to mitigate this constraint would be to devise a way for near real-time detection of such transient phenomena. 
 The \athena/Wide Field Imager (WFI), with its $40' \times 40'$ field of view and large effective area, will detect a large number of X-ray variable or transient objects daily.
 In this work, we discuss an algorithm for the rapid on-board or ground-based detection of X-ray transients with WFI. We present a feasibility test of the algorithm using simulated \athena/WFI data and show that a fairly simple algorithm can effectively detect transient and variable sources in typical \Athena/WFI observations.
\end{abstract}


{\noindent \footnotesize\textbf{*} \linkable{pragati@mit.edu} }


\section{Introduction}
\label{sect:intro}  

A wide variety of astronomical phenomena characterized by changes in flux and spectrum are seen at all cosmological distances and at all time scales ranging from fractions of a second to decades. In the high energy part of the spectrum (X-rays, $\gamma$-rays), this variability  is seen in objects ranging from nearby Galactic compact objects to galactic nuclei at large redshifts. X-ray observatories in space are constantly collecting scientifically interesting  information on these variable and transient sources, which is stored in data archives. With a few exceptions (such as the BAT instrument on the {\it Swift} satellite and the MAXI instrument on the ISS), long delays --- often days to weeks or even months --- occur before transient objects are discovered in ground analysis, by which time the transient event has often died down and cannot be investigated in detail at other wavelengths. 

The launch of X-ray spacecraft like Fermi \cite{Atwood2009THEMISSION}, MAXI \cite{Matsuoka2009TheImages} and \swift/BAT~\cite{Gehrels2004TheMissionc}, which survey large regions of the sky, has been a huge asset to facilitate near real-time reporting of such transient events. The BAT instrument, in particular, detects bright transient sources on-board and transmits key data to the ground within seconds.  Once a time-critical event is acquired from these facilities, there are formal networks like ATEL (The Astronomer's Telegram) \cite{Rutledge1998TheCommunityc}, AMON (Astrophysical Multimessenger Observatory Network) \cite{Smith2012TheAMONc} and the Gamma-ray Coordinates Network / Transient Astronomy Network (GCN/TAN)  \cite{Barthelmy2013GCN/TANFunctionality} that help disseminate the information to a wider astronomical audience for their follow-up.  There have also been serendipitous discoveries of these transient phenomena from 
X-ray spacecraft like 
\chandra, \xmm\ or Ginga, which often get lost in archival data and are often discovered weeks or months later. 
Examples include the discovery of a peculiar flaring source J1806–27 in NGC 6540 \citep{Mereghetti2018} and flares from ISO-Oph 85 \citep{Pizzocaro2016}, both discovered more than a decade after they were first observed by\xmm. Such discoveries are made possible by dedicated efforts like the EXTraS project \citep{luca2017}, aimed specifically at searching the \xmm~archive for transient phenomenon.

In the 2030s, \xmm~ will be succeeded by the Advanced Telescope for High-ENergy Astrophysics \cite[\athena;][]{Nandra2013TheMission} as ESA's large space observatory for the exploration of the X-ray sky. Composed of a microcalorimeter (X-ray Integral Field Unit, X-IFU \cite{Barret2016TheX-IFU}) for  imaging X-ray spectroscopy with high spectral resolution, and a wide field ($40' \times 40'$) X-ray survey instrument (Wide Field Imager, WFI \cite{Meidinger2017TheAthena}), \athena\ will revolutionize the field of X-ray astronomy. Especially important for the sake of X-ray transient science is the surveying capability of WFI, which we briefly mention below. 

The WFI consists of two independently operated detectors, a large detector array and a separate small `fast' detector, with both detectors based on DEPFET active pixel sensor technology.
The energy range of operation is 0.2 - 15 keV with a spectral resolution of $<$ 170 eV at 7 keV. The large detector array consists of 4 detectors of 512 $\times$ 512 pixels spanning the $40' \times 40'$ field of view (FOV), while the small detector is 64 $\times$ 64 pixels and will be operated out of focus, i.e. without imaging capability, to minimize pile-up and optimize throughput performance for bright point sources. 
%
The WFI is thus designed to provide good surveying capability because of the wide field of view and large grasp for performing wide area surveys, low pile-up for bright sources, and absolute temperature and density calibration for in-depth studies of the outskirts of nearby clusters of galaxies. It also has high count-rate capability paired with good spectral resolution, for detailed explorations of bright Galactic compact objects \cite{Rau2016AthenaDrivers} and excellent sensitivity for low luminosity AGNs at high redshifts. 

In order to maximize the science gains for transient science, we have proposed to  contribute a Transient Analysis Module (TAM), to the WFI instrument. The TAM is a software module that could perform on-board transient source detection from the real-time detector data stream. Alternatively, a similar algorithm could perform rapid transient detection in the ground data processing pipeline.

In this paper, we discuss the science enhancement achievable with the TAM and a simple baseline algorithm for on-board transient detection. The paper is organized as follows. We discuss the science benefits in \S\ref{sec: science} and outline the algorithm in \S\ref{sec: algorithm} with proof of concept in \S\ref{sec: poc}, followed by a summary in section \S\ref{sec: summary}.

\section{Wide Field Imager transient science}
\label{sec: science}

The sensitivity limit of \athena/WFI is as low as 1$\times$10$^{-17}$ erg cm$^{-2}$ s$^{-1}$,  and the effective area is $\sim$ 30 times that of \chandra.  
 We find that SIXTE simulations of 
the Chandra Deep Field South field yield $\ge$ 3000 sources per Ms pointing 
(see \S\ref{sec: poc}) as compared to $\sim$ 100-200 with \chandra~and \xmm \cite[]{ranali2013,Luo_2016,barret2013}. 
We can estimate the number of variable sources using \swift\ data: the \swift/XRT Serendipitous Source catalog records $\sim$28,000 variable sources over 8 years \cite{Evans20131SXPS:SPECTRA}, yielding a detection rate of $\sim$ 10$/$day at a median sensitivity of 3 $\times 10^{-13}$ erg cm$^{-2}$ s$^{-1}$. 
\athena/WFI has $\sim$ 100 $\times$ the sensitivity and $\sim$ 3 $\times$ larger FOV compared to 
\swift/XRT; a simple scaling of log(N)-log(S) and FOV suggests thousands per day in the WFI.

In this section, we elaborate on the science benefits achievable by rapid transient detection with WFI. We begin with the science enhancement of known transients science in \S\ref{sub sec: ku}, while the next section expands on the prospects of unknown transients in \S\ref{sub sec: nu}, followed by a section on discovery science in \S\ref{sub sec: uu}. \\

\subsection{Known transients}
\label{sub sec: ku} 
Among a wide variety of X-ray transients, we provide in this section estimates of the probability of {\it serendipitous} detection of such transients in each class, taking one source as a representative for each class. We begin by assuming that a total of 100 counts in some characteristic time scale ($\delta$t) is considered a `detection', the characteristic time scale being defined as the time during which the transient is flaring/variable.  We choose such a large `detection' threshold because we wish to obtain some information on the spectrum and variability of each source.  Assuming a Crab spectrum, the fluence ($\mathcal{F}$) corresponding to 100 WFI counts is 2$\times10^{-11}$ erg cm$^{-2}$. 
The flux (F) to get 100 counts in $\delta$t\,s for each source is therefore (2$\times10^{-11}$)/$\delta$t erg cm$^{-2}$ s$^{-1}$. 

For each source, we first note the flux measured by present (or past) X-ray instruments, F0, and the flux F required for the WFI to attain 100 counts in a characteristic time. The ratio F0/F then gives us an improvement in luminosity distance, I = $\sqrt{\frac{(1+\znew)}{(1+z)}}$, where $z$ is the source redshift and $\znew$ is the redshift limit at which this source would produce 100 WFI counts in $\delta$t s. We then calculate the enclosed volume V (comoving volume at redshift $\znew$) corresponding to  $\znew$\footnote{http://www.astro.ucla.edu/~wright/CosmoCalc.html}. (For our rate estimates, we take the upper limit of V for the observable Universe to be $\znew=11$.  For some extremely bright sources, the number of counts at this limiting redshift exceeds 100.) Thereafter, assuming a conservative WFI duty-cycle ($f_{WFI}$) of 40\% (Arne Rau, private comm.), we proceed with the source rate calculation in one of the following ways: 

\begin{itemize}

\item If the number of sources per galaxy, $n_{src}$, for each source class is available in the literature, we  adopt a slightly different approach. We calculate the number of galaxies, $n$, inside the enclosed volume V ($n$=$\eta\times{V}$, where $\eta$ is the number density of such galaxies). This is then multiplied by the number of such sources per galaxy ($n_{src}$), and finally scaled to the FOV of 40$'$ $\times$ 40$'$ (0.000136\,sr) of WFI, outburst rate per unit time ($r$) and the WFI duty cycle $f_{WFI}$.
Therefore, the number of variable sources detected per year (R$_{D\it{WFI}}$) with WFI is given in Eqn \ref{eq1} below.
\begin{equation}
 R_{D\it{WFI}} = n\times{n_{src}}\times0.000136\times{r}\times{f_{WFI}}/(4\pi)
\label{eq1}
 \end{equation}

\item If the number density (per unit volume) of such sources is available in the literature as R$_{D\it{4\pi}}$, we use that to find the number of such sources in the enclosed volume V. 
We then scale the corresponding surface density to the WFI FOV. Since we are interested in finding the variability with \athena, we multiply this number by the outburst rate of the source per unit time ($r$) 
and the WFI duty cycle $f_{WFI}$. This complete equation that gives the number of sources detected per year (R$_{D\it{WFI}}$) is now as follows 
(Eqn \ref{eq2}).

\begin{equation}
 R_{D\it{WFI}} = R_{D\it{4\pi}}\times{r}\times{V}\times0.000136\times{f_{WFI}}/(4\pi)
 \label{eq2}
\end{equation}

 
 \item In those cases like \emph{Chandra} Fast X-ray Transients where we do not have an estimate of the number density of sources, we proceed in a third way. We use the all-sky event rate at a certain flux level and assume a Euclidean log N-log S relationship, 
 
 \begin{equation}
      N (> S) = N_{0} S^{-\alpha}
      \label{eq3}
 \end{equation}
 
to estimate the number of such events at the WFI flux limit. We then scale this number to the WFI FOV and duty cycle $f_{WFI}$ to obtain an estimate of the number of such events detectable above S, with WFI per year.

 \end{itemize}
 
We note that the source rates throughout the paper do not account for cosmological effects, including corrections for spectral redshifts (k-corrections), cosmic time dilation, or source evolution over cosmic time-scales. Our purpose is to make rough estimates of the probability of detecting sources of different classes, but we leave such corrections to a more detailed analysis. 

\begin{table}
 \centering
 \scriptsize
\caption{ \scriptsize{ The table summarizes the probability of detecting different types of X-ray transients with \athena/WFI, using known transients as exemplars. We have assumed a galaxy density ($\eta$) of 4$\times10^{-3}$ per Mpc$^{-3}$, H$_0$ = 75.0, $\Omega_M$ = 0.30, $\Omega_{vac}$ =
0.70, and a WFI duty cycle ($f_{WFI}$) of 40\%.  The first line for each source provides the source name, characteristic timescale of flares, $\delta T$, the fluence $\mathcal{F}$ and flux $F0$ of a typical observation, applicable parameters ($r$, $n_{\rm src}$, $R_{\rm D\it{4\pi}}$, $\alpha$, $N_{0}$). The second line for each source gives the limiting fluence of $2\times 10^{-11}$ erg cm$^{-2}$ and flux detectable by WFI for 100 counts in $\delta T$; $I$ as the improvement in luminosity distance with $I = \sqrt{\frac{(1+\znew)}{(1+z)}}$, where $z$ is the current redshift of each source and $\znew$ would be achievable with WFI (L indicates local universe); enclosed volume (V), where applicable  and the rate of detection with WFI per year. The equations used for source rate calculation and the source rates are shown in the last column. See \S\ref{sub sec: ku} for details. }}
\vspace{0.5cm}
  \renewcommand{\tabcolsep}{1.7pt}
 \begin{tabular}{cccccccccccc} 

SRC & $\delta T$ & Ref. & $\mathcal{F}$ & Flux & $I$    &  &  & Ref. & & $\znew$  & $R_{\rm D\it{WFI}}$  \\
 & s &  & erg cm$^{-2}$ & erg cm$^{-2}$ s$^{-1}$ &  &  &  &  &   & &  yr$^{-1}$  \\
\hline
\hline
 &&&&&&$r$ &  & & $n_{\rm src}$ & & Using Eqn. \ref{eq1} \\ \\
Magnetar  & 0.040 & \citenum{Kouveliotou1993Recurrent1900+14} & 4$\times10^{-8}$ &  1$\times10^{-6}$&  & 2$^{a}$ yr$^{-1}$ &  & \citenum{ducci2014} & 40 & &  \\
SGR 1900+14 &   & & 2$\times 10^{-11}$ & 5$\times 10^{-10}$ & 45 &  &  &  &   & L& $<$ 1 \\ \\ 

SFXT  & 200 & \citenum{Romano2015GiantDisc} & 2$\times10^{-5}$  & 1$\times10^{-7}$ &  & 2$^{a}$ yr$^{-1}$ & & \citenum{ducci2014} & 40 &  &  \\
IGR J17544-2619 & & & 2$\times10^{-11}$ & 1$\times10^{-13}$ & 1000 &  &  &  &  & L & $<$ 1 \\ \\

\hline 
 &  & &  &  &  & $r * R_{\rm D\it{4\pi}}$ &  & & V & & Using Eqn. \ref{eq2}    \\ 
 & & & & & & & & & Gpc$^{3}$  \\ \\
 
ULX M82 & 100000 & \citenum{Bachetti2014AnStar} & 1$\times10^{-6}$ & 1$\times10^{-11}$ &  & 0.0175 & & \citenum{Swartz_2011,kaaret2006} &  &  & \\
& & & 2$\times10^{-11}$ & 2$\times10^{-16}$ & 224 &  Mpc$^{-3}$ yr$^{-1}$ &  & & & L &$<$ 1 \\ \\

 jetted TDE  & 100000 & \citenum{Burrows2011RelativisticHole} & 2$\times10^{-5}$ & 2$\times10^{-10}$  &  & 0.03  &  & \citenum{Sun_2015} &  & & \\
SW J1644 +573 & & & 2$\times10^{-11}$ & 2$\times10^{-16}$ & 1000 & Gpc$^{-3}$ yr$^{-1}$ & & &  3048 & 11 & $<$ 1\\ \\
 
`faint' TDE  & 100000 & \citenum{esquej2008} & 3$\times10^{-8}$ & 3$\times10^{-13}$  &  & 1$\times10^{5}$  &  & \citenum{Sun_2015} & &  & \\
NGC 3599 & & & 2$\times10^{-11}$ & 2$\times10^{-16}$ &  40 & Gpc$^{-3}$ yr$^{-1}$ & & & & L & $<$ 1  \\ \\

`bright' TDE  & 100000 & \citenum{saxton2012} & 1$\times10^{-7}$ & 1$\times10^{-12}$  &   & 1$\times10^{4}$  &  & \citenum{Sun_2015} &  &  & \\
SDSS J120136.02+300305.5 & & & 2$\times10^{-11}$ & 2$\times10^{-16}$ & 71 & Gpc$^{-3}$ yr$^{-1}$ & & &  1700 & 5 &  75  \\ \\


 Supernova  & 200 & \citenum{snr_time} & 14$\times10^{-8}$ & 7$\times10^{-10}$ &  & 10$^{-4}$  & & \citenum{Taylor2014} & & &  \\
2008d & & & 2$\times10^{-11}$ & 1$\times10^{-13}$ & 85 & Mpc$^{-3}$ yr$^{-1}$ & & & 23  & 0.5 & 10  \\  \\

GRB 060124 & 100 & \citenum{Romano2006PanchromaticAfterglow} & 7$\times10^{-6}$ & 7$\times10^{-8}$  &  & 100-1800 &  & \citenum{Guetta_2007} &  & &  \\
& & & 2$\times 10^{-11}$ & 2$\times 10^{-13}$ & 590  &  Gpc$^{-3}$ yr$^{-1}$ &  & & 3048  &  11 &  1-24  \\ \\

FSRQ  & 100000 & \citenum{Hayashida2015RAPIDOBSERVATIONS} & 12$\times10^{-7}$ & 12$\times10^{-12}$ & & 4*10$^{-7}$ &  & \citenum{ajello2014} & & &  \\
3C 279 & & & 2$\times10^{-11}$ & 2$\times10^{-16}$ & 245  &  Mpc$^{-3}$ yr$^{-1}$ &  & & 3048 & 11 &   5   \\ \\

BL Lacertae  & 40000 & \citenum{Takahashi1996ITALASCA/ITAL421} & 1.5$\times10^{-5}$ & 3.7$\times10^{-10}$ & &  2*10$^{-7}$  & &\citenum{ajello2014} & & &    \\
Mrk 421 & & & 2$\times10^{-11}$ & 5$\times10^{-16}$ & 860 &  Mpc$^{-3}$ yr$^{-1}$ & & & 3048 & 11 &  3  \\ \\

FRB 121102 & 3.2$^{b}$  & \citenum{Scholz2017Simultaneous121102} &  1$\times10^{-9}$ & 3$\times10^{-10}$ &  &  10$^{8}$ & & \citenum{scholz2016,nicholl2017} &  &  &  \\
& & &2$\times10^{-11}$ & 6$\times10^{-12}$ &  7 &   Gpc$^{-3}$ day$^{-1}$& &   &  123 & 1 &  $>$ 1000  \\ \\

 CDF-S XT2 & 100 & \citenum{Xue2019AMerger} & 6$\times10^{-11}$ & 6$\times10^{-13}$ &  &  1.8$\times10^{3}$  & & \citenum{Xue2019AMerger} &  & & \\ 
 (or NS-NS merger)& & & 2$\times10^{-11}$ & 2$\times10^{-13}$ & 2 &  Gpc$^{-3}$ yr$^{-1}$ & & & 214 & 1.3 & 2   \\ \\

\hline 

 &  & &  &  S &   &  $\alpha$  &   &   & N$_{0}$  & &   Using Eqn. \ref{eq3}   \\ \\

XRT110103 & 10 & \citenum{Glennie2015TwoData} & 8.7$\times10^{-10}$  & 8.7$\times10^{-11}$ &   & 3/2 & & \citenum{Glennie2015TwoData} & 4$\times10^{-10}$  & & 2  \\ 
 & & & 2$\times10^{-11}$ & 2$\times10^{-12}$ & 7 &  &    &  & yr$^{-1}$ 
 & & 578 \\ \\


\hline 







\hline 

\\

\label{source_rate}

\end{tabular} 

$^{a}$ Typical values \\
$^{b}$\chandra~readout time during the observation \\



\end{table}


We now go on to introduce each class of sources, taking one representative example of each class, and also discuss the probability of detecting the variability in each one of them. We also estimate the improvement in depth achievable with WFI. These values are tabulated in Table \ref{source_rate}. 

\begin{itemize}
    \item \textbf{Galactic compact objects:} 
    The Galactic compact objects comprise white dwarfs, neutron stars and stellar black holes, either in binary systems or in isolation. They often exhibit variability: for instance, owing to changes in accretion, thermonuclear burning causing Type I/II bursts,  state changes in black holes, or magnetar flashes. The time-scales are varied, ranging from milliseconds to a decade. For instance, in the case of the magnetar SGR 1900+14, BATSE detected outbursts with a characteristic time scale ($\delta$t) of 40 ms \cite{Kouveliotou1993Recurrent1900+14}. The probability of serendipitously detecting magnetars like SGR 1900+14 yielding 100 counts in this 40 ms is inevitably very low. However, the important consideration is the improvement in detection depth of 45 times with WFI compared to its actual distance, since WFI can detect a flux that is only 0.05\% of its actual flux, allowing us to probe deeper for magnetars than any other X-ray spacecraft so far. Similarly for the case of the supergiant fast X-ray transient (SFXT) IGR J17544-2619 (known for quiescent emission most of the time interrupted by sudden random flares),
WFI could detect this object 1000 times farther away, delving into a new discovery space. 

      \item  \textbf{Ultraluminous X-ray sources (ULXs):}
    These X-ray sources have Eddington luminosities larger than that of stellar mass objects, ranging around 10$^{39}$ erg s$^{-1}$. These are thought to result from beamed emission from X-ray binary systems containing a heavy neutron star or an intermediate black hole. The variability in ULXs is often attributed to changes in the accretion rate and lasts from a few ks to years \cite{Swartz2004TheGalaxies,King2016ULXs:Holes}. Taking ULX M82 \cite{Bachetti2014AnStar} as a typical example, we would be able to detect 100 counts in 100 ks at a distance of $\sim$ 224$\times$ farther than ULX M82 in the local Universe. 
   
    A significant fraction of ULXs exhibit transient pulsations that are often detected during their `high' state, possibly because of low counts captured during normal flux states. The excellent sensitivity of \athena~
   will enable pulsation searches as well as investigation of propeller effects in ULXs, even in their normal flux states \cite{song2020}.
   
    \item \textbf{Tidal disruption events (TDEs):} TDEs \cite{Thorp2018TidalSurveys} occur when the tidal forces exerted on a star upon close approach to a massive black hole overcome its self gravity and pull it apart.  If we consider 100 counts being detected in 100 ks in a jetted TDE like Swift
    J1644+573  \cite{Burrows2011RelativisticHole}, we will detect these events across the whole observable universe. `Normal' non-jetted TDEs like NGC 3599 and SDSS J120136.02+300305.5 will be detected to significant cosmological distances, up to redshifts of 5 for the latter. If we take SDSS J120136.02+300305.5 as being representative, we estimate that WFI would detect $\sim$ 75 per year. 
    
       \item \textbf{Core collapse supernovae (CCSn):} Core collapse supernovae are spectacular explosions that mark the violent deaths of massive stars. These events are the most energetic explosions in the cosmos, releasing energy of order 10$^{51}$ ergs. The supernova shock breakout lasts for hundreds of seconds. For CCSn like SN2008D with $\delta$t of $\sim$ 200 s \cite{snr_time}, we will detect $\sim$ 10 per year and up to 85$\times$ farther than SN2008D.
    
    \item \textbf{Gamma-ray bursts (GRBs): }
    GRBs \cite{Levan2016Gamma-rayProgenitorsc} are extremely energetic explosions that can last from ten milliseconds to several hours. If the event is less (more) than 2s, it is termed a short (long) GRB. After the initial flash in gamma rays, a longer-lived afterglow --- lasting typically from hundreds of seconds to days or weeks ---   is  emitted at longer wavelengths \cite{Zhang2007Gamma-rayAfterglows}. In the case of the long GRB 060124, where the \swift\ BAT triggered on a precursor and the \swift\ XRT therefore measured the entire X-ray light curve of the prompt emission, the brightest part of the X-ray emission lasted for $\sim$ 100 s \cite{Romano2006PanchromaticAfterglow}.  Considering the improvement in the luminosity distance, WFI could detect this burst 
    even at the farthest redshift of $\znew=11$. We expect to serendipitously detect $\sim$  1-24 GRBs per year with the WFI.

    \begin{figure}
  \centering
\includegraphics[scale=0.75,angle=0]{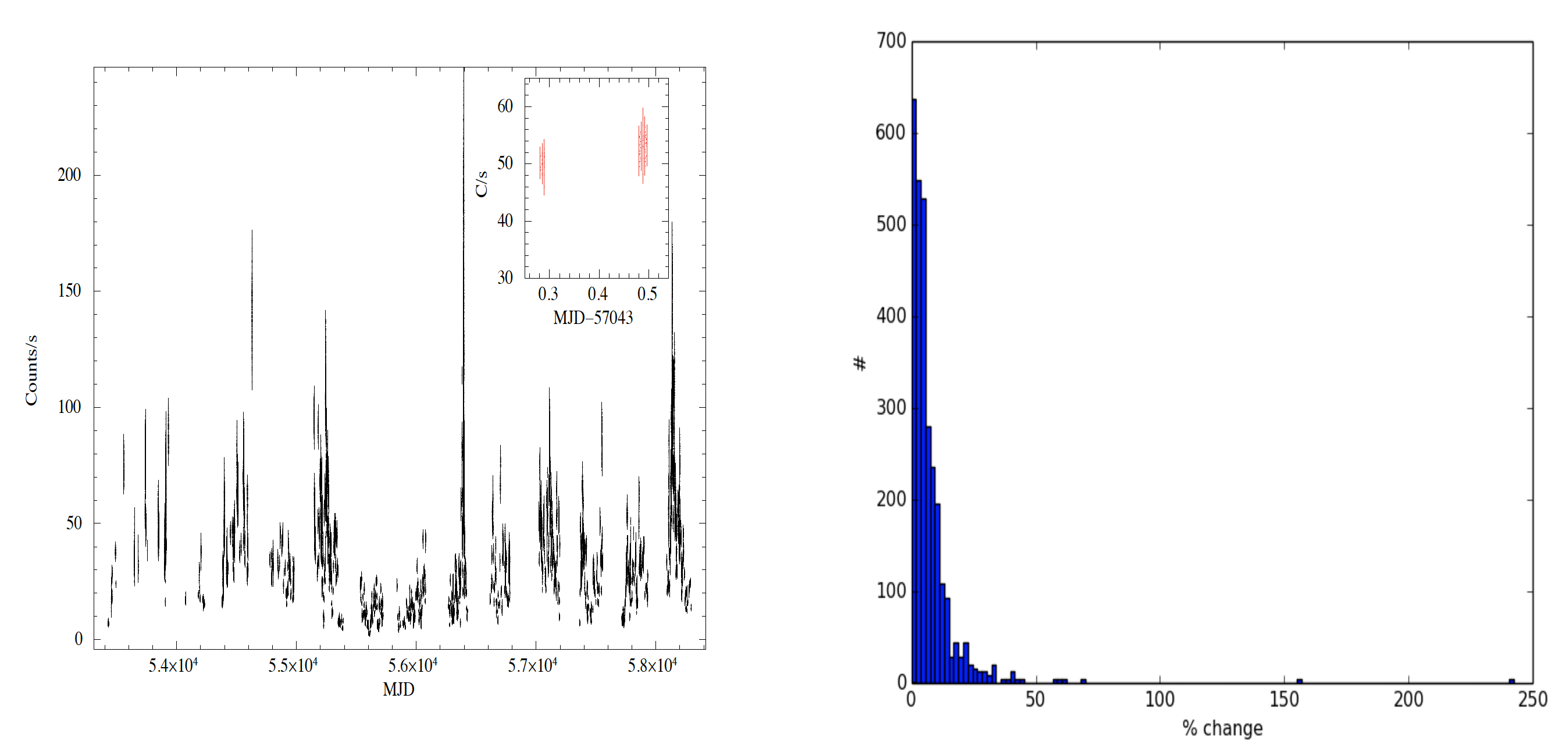}
\caption 
{ \label{mrk_var}
Left: Long-term \swift/BAT light curve of Mrk 421 with 4-hour integration times per data point. The inset shows two random time segments between which variability is calculated. Right: Histogram of the percentage change in count rate for the \swift/BAT light curve of Mrk 421 between successive data points.} 
\end{figure}

 \item \textbf{Active galactic nuclei (AGN):} An AGN \cite{Fabian1999ActiveNuclei.} is a compact region at the center of a galaxy that is very luminous, emitting  10$^{40-47}$ erg s$^{-1}$. The radiation from an AGN is believed to result from the accretion of matter by a supermassive black hole at the center of its host `active' galaxy \cite{Akiyama2019FirstHole}. Three important classes of AGN are: (i) the Seyfert galaxies, which have modest luminosities but are best studied since they are relatively close; (ii) the quasars, which are more luminous than the host galaxy and are particularly numerous at a redshift of $\sim$ 2; and (iii) the blazars, including BL Lacs as well as FSRQs (flat spectrum radio quasars) and OVVs (optically violent variables), seen when our line of sight lies close to the direction of a jet. The X-ray variability in AGNs, caused either by accretion rate or environment change or jets, lasts from $\sim$ minutes to $\sim$ months.
 In the case of blazars, if 100 counts are to be detected in a typical 100\,ks observation, we will detect $\sim$ 5 such flares from blazars similar to the FSRQ 3C 279, or $\sim$3 from Mrk 421 -like BL Lacs, anywhere in the observable Universe.  Of course, different thresholds of detection and different variability timescale probes would produce different rates of blazar flare detection.
 Although there have been extensive studies of AGN variability,  new discoveries continue to intrigue the astrophysics community. For instance, X-ray variability characterized by short, high-amplitude, quasi-periodic X-ray bursts over a rather stable baseline flux, termed quasi-periodic eruptions (QPEs), was recently observed for the first time in GSN 069 \cite{Miniutti2019}.
 Such variability, possibly caused by instabilities of the accretion flow, is reminiscence of `heartbeat' oscillations seen in BH binaries like, GRS 1915+105 \cite{neilsen2011} and can explored by \athena~ to much lower flux levels than current observatories. 


   The TAM algorithm will detect variability on time scales of few kiloseconds  or less. To illustrate the utility of this capability for studying AGN, we investigate the variability on timescales of ks in the long-term \swift~BAT observations of Mrk 421. The \swift\ light curve (left panel of Fig.~\ref{mrk_var}) was divided into segments spanning 4 hours. 
    We then calculated the percentage of variability change in the weighted average in one observation to the next. A histogram of the percentage of variability is shown on the right of Fig.~\ref{mrk_var}. The figure shows that there are hundreds of such cases where the percentage change over 4 hours is significant. 

     In addition to the compact region X-ray variability discussed so far, non-compact region variability from nearby luminous galaxies subtending at least tens of arcseconds can also be detected. However, the  5 arcsecond PSF of the \athena~mirrors will not be sufficient to explore  X-ray lensing of quasars and galaxies, which require sub-arcsecond spatial resolution. 

     \item  \textbf{Fast Radio Bursts (FRBs):}
    FRBs \cite{Lorimer2007AOrigin} are transient pulses discovered in the radio band, characterized by large dispersion measure and timescales of milliseconds. Their origins are unclear; possible explanations include giant SGR flares or coalescing compact objects (for example, see Reference \citenum{Champion2016FiveBursts}). While most FRBs are transient in nature, at least two  (FRB 121102 and FRB 180814.J0422+73)  have repeated outbursts \cite{Scholz2017Simultaneous121102,Collaboration2019ABursts}. For typical event rates of FRBs as 10$^{4}$ sky$^{-1}$ day$^{-1}$ \cite{scholz2016}, and FRB distribution of 10$^{4}$ Gpc$^{-3}$ \cite{nicholl2017} and assuming that the typical X-ray flux is equal to the upper limit found for FRB 121102 during its \chandra~observation \cite{scholz2016}, we could detect more than 1000 such FRBs per year with WFI. We however caution that since no FRBs have yet been detected in X-rays\footnote{If X-ray bursts from SGR 1935+2154 is associated with an FRB \cite{zhang2020}, its source detection rate  and improvement in distance, using X-ray flux and $\delta T$ from NICER \cite{younes2020}, is of similar order for the magnetar SGR 1900+14 derived here.}, this number is highly uncertain.



\end{itemize}

\subsection{Unknown transients}
\label{sub sec: nu}

\begin{figure}
  \centering
  \includegraphics[scale=0.8,angle=0]{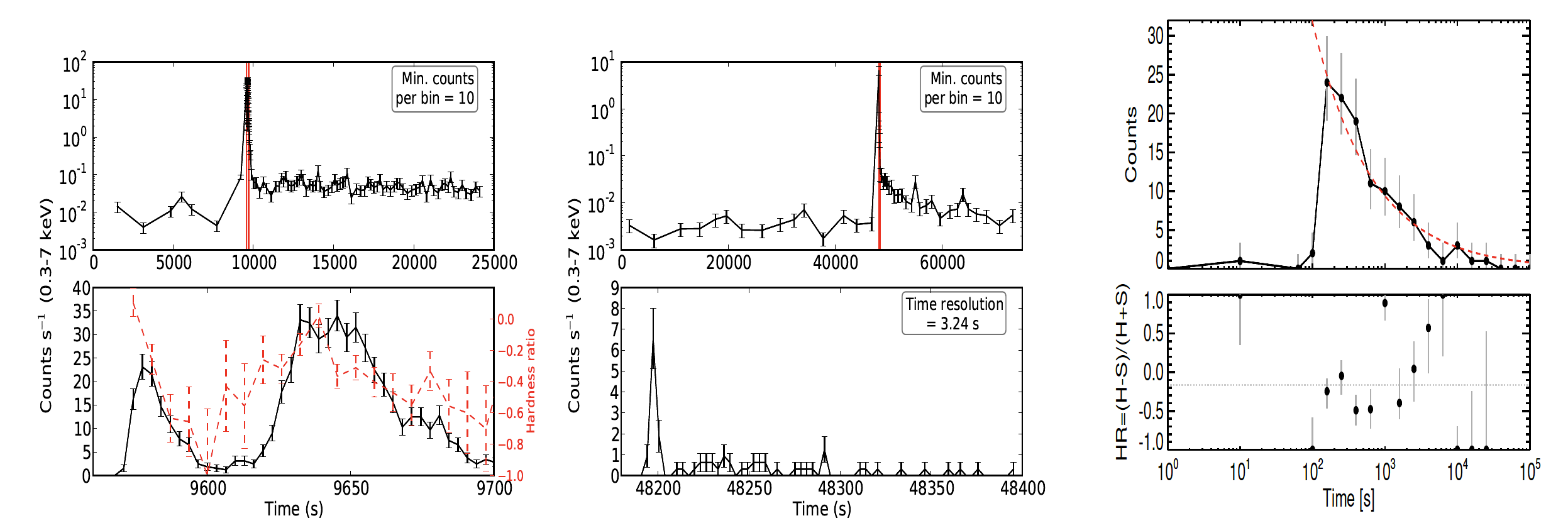}
\caption 
{ \label{chandra_lc}
Light curves for different \chandra\ transients. Left: XRT 000519 \cite{Jonker2013DiscoveryM86}; Center: XRT 120830 \cite{Glennie2015TwoData}; Right: CDF-S XT1 \cite{Bauer2017ATransients}. The variability behavior of such transients are currently unexplained by any known transient behavior. See text for details.} 
\end{figure}

A remarkable new type of fast X-ray transients has been found in \chandra~data. These exhibit X-ray and multi-wavelength properties unfamiliar to known X-ray transients discussed in section \ref{sub sec: ku}. The first detection, named XRT 000519 (Fig.~\ref{chandra_lc}, left), showed a double-peaked light curve \cite{Jonker2013DiscoveryM86}. 
This transient was found close to M86 in the Virgo cluster and the flux increased from being undetected to a peak \chandra~count rate of 20 counts s$^{-1}$ in 10\,s, decayed gradually in 20\,s to $\sim$ 1 count s$^{-1}$, and rose again to count rate of 24 counts s$^{-1}$. The second peak had a flat top lasting around 20\,s before gradually decreasing on a timescale of about 100\,s, followed by a power-law decay with index of $\sim$ 0.3 for 20 \,ks until the observation ended \cite{Jonker2013DiscoveryM86}. Possible mechanisms include the disruption of a compact white dwarf star by an intermediate black hole, but alternative scenarios like a foreground neutron star accreting an asteroid or an off-axis (short) $\gamma$-ray burst are also possible. Similar light curve behavior was also found for XRT 110103, although it did not exhibit a precursor nor a twin peak in the main flare and was a factor of a few fainter compared to XRT 000519 \cite{Glennie2015TwoData}. With an all sky distribution of $1.4 \times10^{5}$  yr$^{-1}$ sky$^{-1}$ with flux $S  > 2 \times 10^{-10}$ erg cm$^{-2}$ s$^{-1}$ \cite{Glennie2015TwoData}, and assuming a log N-log S distribution, N ($>$ S) = N$_{0}$ S$^{-\alpha}$,  with slope $\alpha= 3/2$, we would detect $\sim 2$ XRT 110103 like events per year with WFI.

Another fast transient is XRT 120830 (Fig.~\ref{chandra_lc}, middle).  The count rate increased by three orders of magnitude from the background level reaching a peak and rapidly decayed by more than an order of magnitude. This rise and decay occurred in approximately 10\,s. The transient continued to decay further with some marginally significant flaring events $\sim$ 7 and 14 ks after the main burst. The flare from XRT 120830 could be an X-ray flare from a late M or early L dwarf star with possible minor flares $\sim$ 7 and 14\,ks after main flaring related to the rotation period of the star \cite{Glennie2015TwoData}. 

The X-ray event named CDF-S XT1 (Fig.~\ref{chandra_lc}, right) produced $\sim$ 115 net counts in \chandra, with a light curve characterized by a $\sim$ 100\,s rise time and a power-law decay time slope of $\sim$ -1.53 \cite{Bauer2017ATransients}. There have been speculations of the origin of its transient behavior as an orphan X-ray afterglow from an off-axis short-duration GRB, a dimmer and farther GRB, or a beamed and less variable TDE wherein an intermediate black hole is engulfing a white dwarf. However, none of the above scenarios can completely explain all observed properties of this X-ray flare \cite{Bauer2017ATransients}. 

Another peculiar X-ray flaring source was found near the galaxy NGC 4697, where two brief, ultraluminous flares (separated by four years) were seen. These flares were characterized by a flux increase of a factor of 90 in about one minute  \cite{Irwin2016UltraluminousGalaxiesc}. Since only two such examples were detected among several thousand X-ray point sources within 70 Chandra observations of nearby galaxies, it is plausible that the Milky Way has no analogs to these sources. Given the small number ($\sim$ 40) of X-ray sources in the Milky Way brighter than 10$^{37}$ erg s$^{-1}$, lack of X-ray binaries more luminous than 10$^{38}$ erg s$^{-1}$ in Galactic globular clusters, and rarity of burst sources in the extragalactic sample, a detection of only two seem right  \cite{Irwin2016UltraluminousGalaxiesc}. The nature of these sources remains largely uncertain, and rapid multi-wavelength follow-up of such detections is probably the \emph{only} way to probe the nature of such erratic transients. 
That requires rapid identification of these sources either on the spacecraft or in rapid ground processing.

We also note that, although the transients discussed above are widely categorized as fast X-ray transients, each type of transient is unique in its variability behavior. The latest discovery is named CDF-S XT2 and is speculated to be an aftermath of a NS-NS merger \cite{Xue2019AMerger}. The probable rate of detecting CDF-S XT2 type events (and potentially similar NS-NS mergers that have similar event rates) with WFI is nearly 2 yr$^{-1}$.


Overall, the CDF-S transients are speculated to be either a part of an unexplored regime for known transient class or a new variable phenomena whose nature is unknown. The potential for discovery science from these on-board triggers are therefore significant. \athena\ will be in a much better position to characterize the light curves in detail and probe fainter (and perhaps more abundant) versions of these transients, and will strongly benefit from rapid multi-wavelength follow-up to help constrain their physical nature. 


\subsection{Discovery space and synergy with other facilities}
\label{sub sec: uu}
As discussed earlier, WFI will have the capability to probe into parameter space not explored by any X-ray observatory to date. This will undoubtedly open new avenues for discovery science, allowing us to delve into an unexplored regime of the X-ray Universe. Many excellent AGN targets for \athena~will be within the LSST sample and spectroscopy of the first quasars from Euclid, WFIRST, and LSST. 

While our main focus of this paper is to demonstrate the usefulness of an on-board or ground-based rapid
transient detection system in \athena/WFI, it is worthwhile to consider some of the multiwavelength studies obtainable by the synergy between \athena~ and other observatories in the 2030s (see e.g., Athena Multi-messenger and High Energy Astrophysics Synergy, by Piro et al). \athena~will be operating in an era of deep multiwavelength extragalactic surveys like the Large Synoptic Survey Telescope (LSST). Given its wide FOV, the LSST is capable of imaging its field of regard 
in 3-4 nights; LSST will therefore observe millions of AGNs during its operation \cite{lsst2019}. 
\athena~observations will 
likely be complementary to LSST, since optical surveys detect very luminous AGNs while \athena~will be useful to uncover the X-ray emission of `fainter' AGNs at high redshifts to constrain the seeds/processes that led to the early growth of SMBHs \cite{rau2016}. 
\athena~and the Square Kilometer Array (SKA) will also have excellent synergy for studying objects in very different energy regimes. While the survey strategy for SKA and WFI may have room for a planned overlap for observing similar regions in sky, one needs to keep in mind that the SKA will typically have better (sub-arcsecond) angular resolution than \athena . Targets will therefore have to be carefully selected taking this into account. There have already been combined efforts to investigate the science enabled by the SKA and \athena~surveys. These include a large range of astrophysical topics, from the very first stars and galaxies to transients at all timescales \citep{cassano2018skaathena}. \\
Important multimessenger synergies also exist between LISA and \athena. LISA will localize gravitational wave emission from any point on the sky; \athena~will be able to observe locations provided by LISA and observe X-rays from the surrounding gas of the newly born black hole. Recent studies have shown that $\sim$ 10 BH binaries in the mass range of 10$^5$ - 10$^8$ M$_{\odot}$ discovered by LISA at z $<$ 3.5 can be detected by \athena~in 100 \,ks, for a prompt X-ray emission of $\sim$ 1 - 10$\%$ of the Eddington luminosity \cite{mcgee2020}. \\
With its unparalleled capabilities, \athena~in the 2030s will therefore be a transformational observatory, operating in tandem with other observatories spanning wide electromagnetic spectrum with SKA, ALMA, ELT, JWST, CTA, etc, and large efforts to plan for this science are already underway \cite{alma2015,padovani2017,takahashi2013}.


\section{Methodology}
\label{sec: method}
The primary aim of the TAM is to  generate 
alerts about transient activity, either for detection of new transients or for variable behavior for known sources. In this section, we will discuss step-wise the algorithm we developed for the case of on-board detection, followed by the proof-of-concept of this algorithm. While it currently appears that this module will not be included in the WFI instrument, a similar capability is under consideration for the WFI ground processing pipeline.  As we show below, this on-board processing is feasible with modern flight computers, and this capability could be considered for other  X-ray telescopes in the future.

\subsection{An outline of the algorithm for on-board transient detection}
\label{sec: algorithm}
\begin{figure}   
 \centering
   \includegraphics[scale=0.42,angle=0]{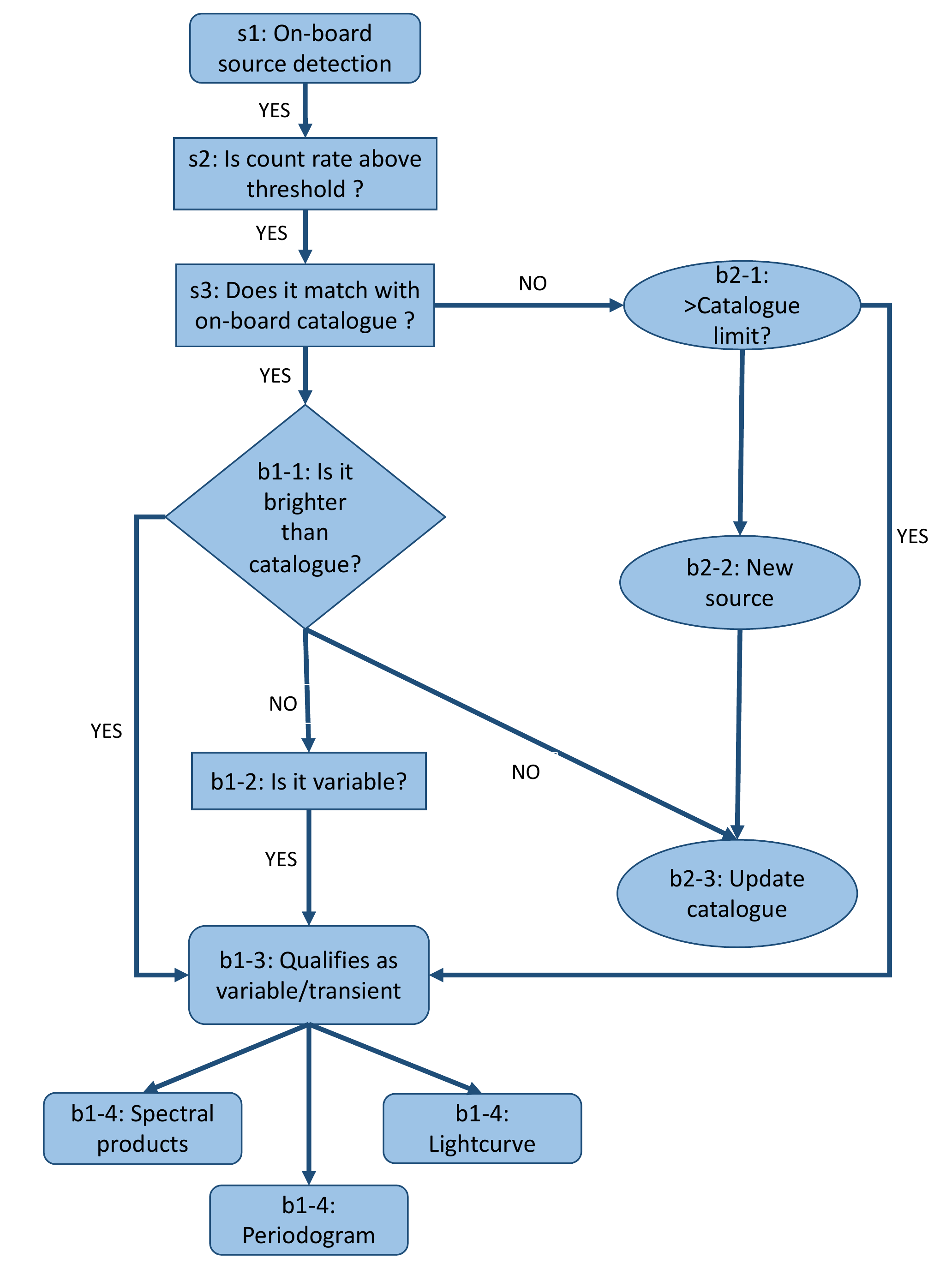}
\caption 
{\label{fc}
Flowchart to illustrate the on-board detection of transients with \athena/WFI. See section \ref{sec: algorithm} for details. } 
\end{figure}

The step-wise implementation of the algorithm is demonstrated by the flow-chart shown in Fig.~\ref{fc}. We emphasize that this proof-of-concept algorithm was developed to show that on-board transient detection is feasible within the computing constraints of the WFI instrument and was tailored to demonstrate compliance with a set of requirements from the WFI team.

The algorithm begins with the detection of transients with the WFI on-board \athena\ (s1 in Fig.~\ref{fc}). We then check in step two (s2) whether the count rate for this detection is above a specified threshold chosen to be 30 counts here.\footnote{We also successfully demonstrated the algorithm efficacy by assuming a threshold of 50 counts as chosen by the WFI team for these tests.} This is important since we would only want to send on-board alerts for the most interesting/bright objects for their follow-up. Depending on whether the source matches with any source position in the on-board catalog or not (s3), the subsequent steps either proceed in the downward vertical direction as branch 1 (b1) or in the horizontal direction as branch 2 (b2). 
If the position of the detected transient matches with the on-board catalog, it is a known source; if it is brighter than the catalog flux (b1-1), or if it is variable during the observation (b1-2), we have detected variability in a known source, and we flag this source (b1-3) and generate science products like light curves in several energy bands, hardness ratios, and a periodogram (b1-4). 
Branch 2 starts if the detected transient does not match with the on-board catalog (i.e., s3).  We then check whether the flux for this detected transient is above the catalog limit (b2-1). If it is, the source is bright and qualifies as a variable/transient (b1-3), which subsequently leads to production of science products in step b1-4. In any case, if the source does not match with any known source, we flag it as new (b2-2) and update the on-board catalog with the information from this new detected source (b2-3). The on-board catalog is also  updated when the flux of a known source is dimmer than its catalog value (b1-1 to b2-3). We note that that the algorithm shown is designed to detect transients that grow more luminous than their normal state. However, the algorithm can easily be modified to also detect source `dimming'.  

The utility of on-board alerts is, however, inevitably dependent on the frequency of the satellite's contact with the ground station. The current mission strategy for \athena\ is a single ground pass of 4 hours per day (Arne Rau, private communication).  While the daily ground pass produces a latency of up to 20 hours for distribution of on-board alerts to the community, the 4-hour window permits real-time alerts with $<$ 1ks latency about 16\% of time. 
These prompt transient alerts will facilitate multi-wavelength observations, thereby allowing us a unique opportunity to investigate these enigmatic objects in many wavelengths simultaneously, as discussed in \S\ref{sec: science}.

\clearpage
\subsection{Testing of algorithm: Proof-of-concept}
\label{sec: poc}
\color{black}
In order to provide a demonstration of the algorithm outlined above, we created simulated WFI data sets of 1~ks duration using SIXTE\footnote{https://www.sternwarte.uni-erlangen.de/research/sixte/}.  We used the Lehmer et al. catalog \cite{Lehmer2005TheCatalogs} of sources from the extended Chandra Deep Field South (CDFS) to simulate a typical field of X-ray background sources, and included the effects of spacecraft dither with the attitude file `CDFS\_lissajous\_80ksec.att' (available on the SIXTE website). To this simulated data set, we added 6 transient sources of varying intensities with their light curves and spectra taken from \xmm~observations, either in the flaring phase or quiescent emission. 

\begin{figure}
  \centering
  \includegraphics[scale=0.75,angle=0]{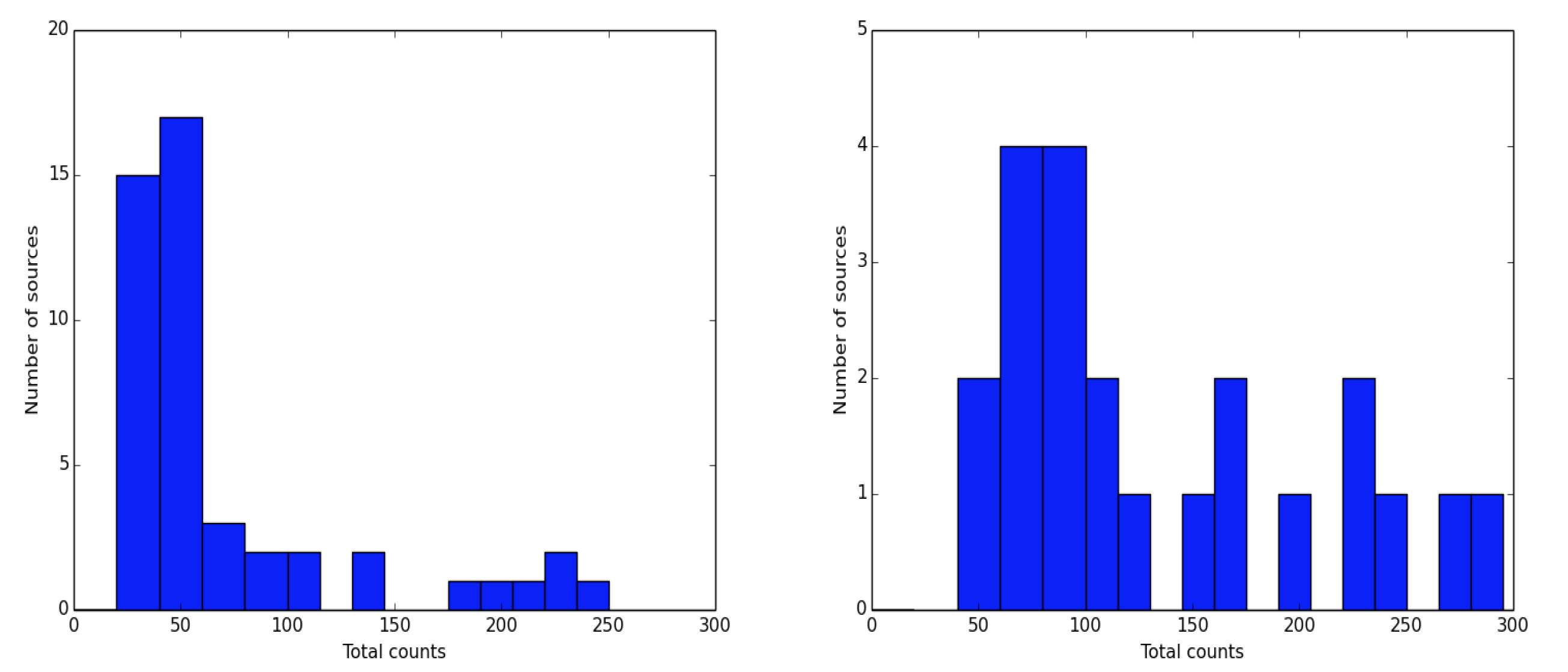}
\caption 
{ \label{histogram}
Left: Histogram for the number of sources detected when only the instrumental background was included. Right: Histogram for the number of sources assuming a high background with $\sim$ 400 counts/ks per source region for the same field. In the right figure even with high background, we still detect bright sources.}
\end{figure}

In the first step of the algorithm, s1 (see Fig.~ \ref{fc}), the algorithm does a blind source detection, using `wavdetect' with a false-positive probability threshold (sigthresh) of 1E$^{-10}$, background significance threshold of 1E$^{-6}$ and wavelet scales of (2.0, 4.0).
We provided a PSF file giving the \athena/WFI PSF size for each image pixel.
For step 2 (s2) we set the threshold for detection to 30 counts. 
Step s3 then compares the position of the detected sources with the ones in the Lehmer catalog. If the angular difference between two a detected source and a catalog source is less than a conservative value of 7 arc-secs, we consider it a match, and the algorithm proceeds  to the step (b1-1) where we compare the count rates between the source and the catalog. If the source rate exceeds the catalog rate, a transient has been detected and we proceed to step b1-4 where the science products (light curve, hardness ratio, and periodogram) are extracted.  The sources detected by this exercise are shown in the left panel of Fig.~\ref{histogram} . 

In order to test the algorithm performance for sources in the direction of a bright background source such as a galaxy cluster or supernova remnant, we also checked how many sources are detected while including a large X-ray background of 400 counts/ks/source region  (each source region being the average of background region obtained from `wavdetect' during source detection, $\sim$ 250 pixels). We have intentionally chosen a high background rate to ensure that the observations were background-dominated, in order to bound the problem in two cases: without/with background. The histogram for source detection with the background is shown in the right panel of Fig.~\ref{histogram}. Even with such a large background, the brighter sources are still detectable by WFI.

\begin{figure}
  \centering
   \includegraphics[height=10cm,width=10cm,angle=0]{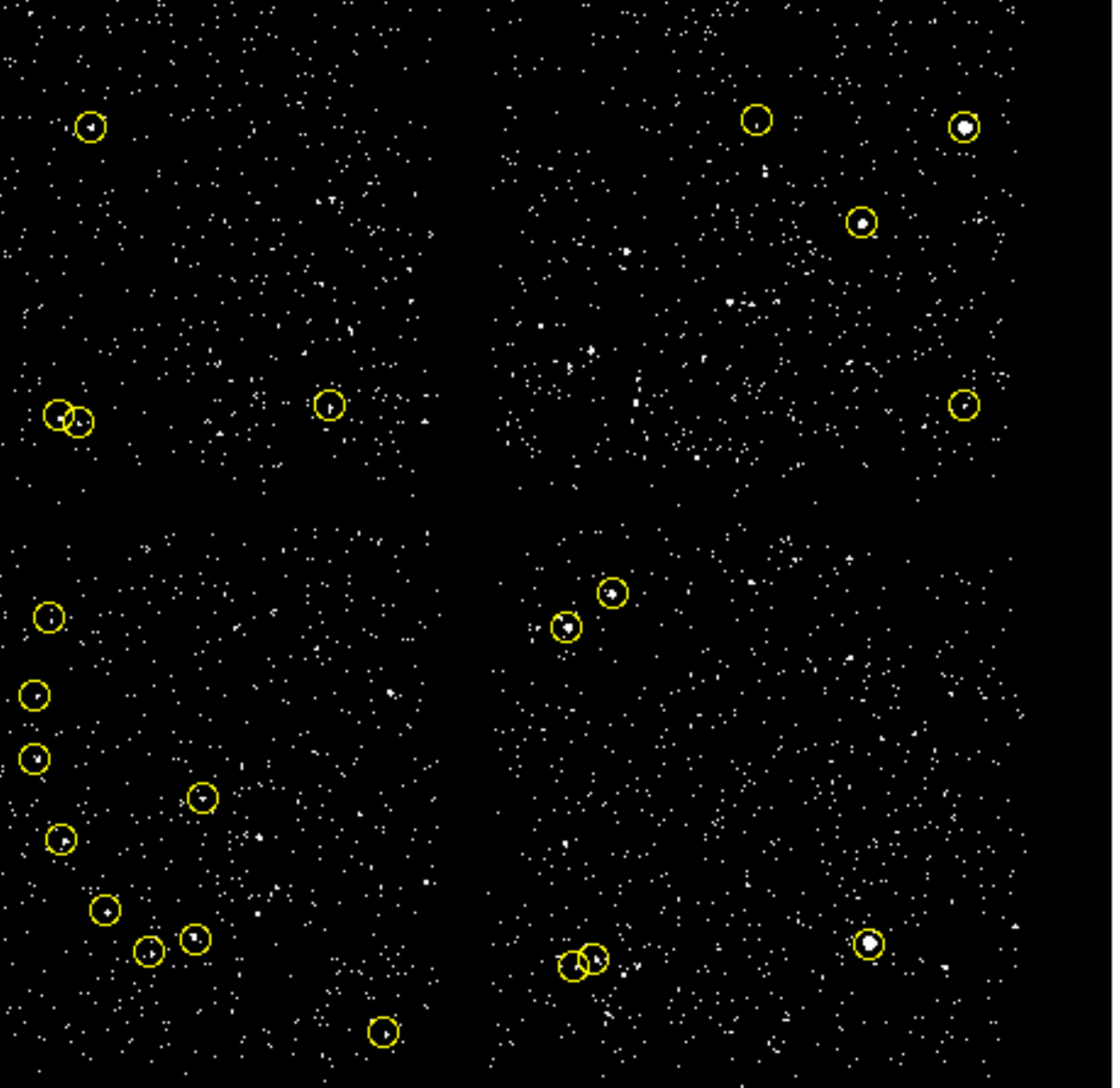}
\caption 
{ \label{fig1}
SIXTE simulated CDFS field for WFI for an exposure of 1\,ks with 22 transients that were inserted (marked with yellow circles). The misalignment of the image arises because the image was created with dither on.} 
\end{figure}

 Next, in order to ensure that we are indeed measuring the variability in detected sources, we inserted 15 more artificial test sources with light curves shaped as Gaussians, square waves, pulses, step functions or sine waves, with each light curve shape having one source each with $\sim$ 30, 50 or 100 counts in total. The spectral shape for all these light curves was assumed to be Crab-like. We also inserted one source with the \xmm~ light curve and spectrum of the pulsar LMC X-4. All 22 sources that were injected in the WFI CDFS image are shown in Fig.~\ref{fig1}.  

\begin{figure}
  \centering
  \includegraphics[scale=0.75,angle=0]{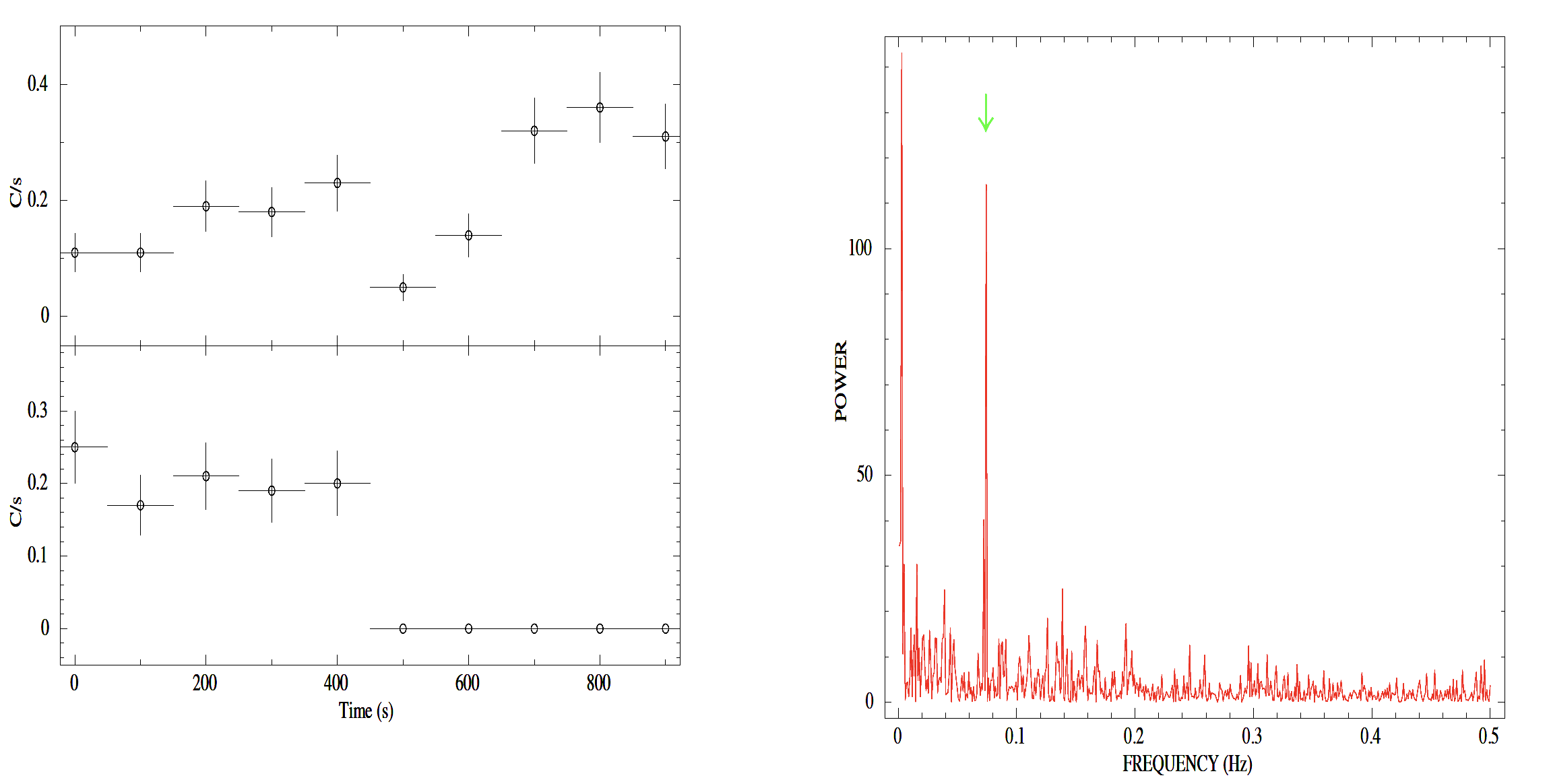}
\caption 
{ \label{lmc_per}
Left: Variability detection for two inserted light curves (upper: a CDFS source; lower: an artificial injected transient with a `step-function' light curve). Right: Periodicity determination of LMC X-4 as detected by the algorithm. The true periodicity of the pulsar is marked with arrow.} 
\end{figure}

The execution of the algorithm showed that we were able to retrieve all the transients already in the catalog as well as the inserted ones. Irrespective of whether the source was brighter than the catalog, we looked for variability in segments of 100\,s. We define variability as a 5\,$\sigma$ deviation of the count rate in each 100 s from the median count rate over the whole 1 ks. For the purposes of this test, we only generated light curves as science products for these 100 s segments. We recovered all such segments where the criteria are satisfied and verified them against the inserted light curves. Two such variable sources are shown on the left of Fig.~\ref{lmc_per}. The upper light curve is from a CDFS transient while the lower one is from an injected transient with a step function light curve. 

We were also able to detect the periodicity for brighter sources with large significance. One such periodogram for an inserted transient (LMC X-4) is shown in Fig.~\ref{lmc_per}. We caution that an artificial periodicity at the readout time of 5\,ms in the `normal' mode and 80\,$\mu$s in the `fast' detector mode will be seen in the periodogram so any periodic detection around those values (and their harmonics) will have to be carefully examined. Note that, in order to generate light curves, we have used the SIXTE command `makelc'. The hardness ratio was calculated as a ratio of count rates in the bands 0.5-2.0 keV and 2.0-15 keV. Periodicities in the light curves were determined using the \texttt{FTOOLS} command `powspec' on the 1 s binned light curves.

In order to mimic the uncertainties arising from star tracker errors, we executed the algorithm 100 times with random positional errors of $\pm$ 3 arcseconds on each axis. We were able to detect 99\% of sources with astrometry errors $<$  7 arcseconds, 
thereby demonstrating the robustness of the on-board detection to spacecraft attitude errors.

Timing tests of the algorithm were executed in order to verify that the target space-qualified CPU could execute the required processing in real time.  The version of the proof-of-concept algorithm used in this test included a combination of shell scripts, SIXTE analysis tools, CIAO code, FTOOLs, and Python code.  Tests were run under Linux on a Dell Inspiron 15-3000 series laptop with an Intel Core i5-5200U CPU running at 2.20 GHz.  We used Dhrystone and Coremark benchmark tests to estimate the execution time ratio between this machine and the flight target dual-core Leon-3FT processor.  Our tests showed that the flight CPU could fully process the simulated data, including production of exposure maps, images, source detection, and variability tests, and output source positions, light curves in two energy bands, harness ratios, and periodograms, in 50\% of real time.  We expect that optimization of the code could significantly improve on this performance.

The discussions throughout the paper assume the execution of the on-board algorithm during normal operating modes of WFI. We also demonstrated the implementation of this algorithm on simulated slew survey data, similar to the highly successful XMM slew survey. Should this be implemented, a rough estimate indicates that with WFI, the possible number of {\it highly variable} sources with flux changes of a factor of ten and more detectable in the slews will be $\sim$ 7 per year. 

To summarize, we have verified that our algorithm detects all types of variable sources, including the inserted sources that were not included in the catalog and were therefore classified as transients.


\section{Summary}
\label{sec: summary}
With sensitivity better than current existing spacecraft like \chandra\ and \xmm, \athena/WFI will enrich our understanding of the known transients and will contribute to the discovery and study of new X-ray transients. This increase in sensitivity coupled with faster readout times will also aid in probing faint X-ray transients, which has so far been hindered by current instruments that have typical readout of seconds. In this paper, we have presented an algorithm for rapid on-board or ground-based detection of X-ray transients with the WFI onboard \athena. In addition to this improvement in the understanding of transients with TAM, such alerts will also facilitate multiwavelength follow-ups with other observatories. 

Tests using simulated data with artificial and real X-ray variable and transient sources show that this simple algorithm can successfully detect both transient and variable sources on time scales of $<$ 1~ks with the available computational resources, and the results could be relayed to the ground with very low latency for the 16\% of the time that \athena\ is in contact with the ground station.   This is expected to produce significant numbers of prompt alerts for interesting transient X-ray sources.  Discussions with NASA and the WFI team are continuing to determine whether this  on-board transient detection capability will be included on \athena.




\section*{Acknowledgements}
This research has made use of data and/or software provided by the High Energy Astrophysics Science Archive Research Center (HEASARC), which is a service of the Astrophysics Science Division at NASA/GSFC. The authors would like to thank the SIXTE team especially Thomas Dauser and J{\"o}rn Wilms for their help with SIXTE simulations. We also thank the anonymous referees for their useful comments and suggestions. 
This work was carried out under NASA grant NNX17AB07G.

\bibliographystyle{spiejour} 
\bibliography{mendeley_v3}

\section*{Biographies}

Pragati Pradhan is currently a post-doctoral associate at MIT prior to which she was a post-doctoral scholar at Penn State University. She obtained her Ph.D from Raman Research Institute/University of North Bengal during which time she also served as an Assistant Professor at St. Joseph’s College, Darjeeling. Her research interests mainly include accreting neutron stars and their transient behavior. She also works on stellar wind diagnostics of massive OB and WR stars with the Chandra/HETG group at MIT. 

Abraham Falcone is a Research Professor of Astronomy and Astrophysics at the Pennsylvania State University. He received his BS (Physics) from Virginia Tech in 1995 and PhD (Physics) from the University of New Hampshire in 2001. He has authored $>$ 200 refereed journal publications. He leads research in fields ranging from X-ray and gamma-ray instrumentation to the study of particle acceleration at high-energy astrophysical sites such as active galactic nuclei, gamma ray bursts, and X-ray binaries.

Dr Jamie A Kennea is the Lead of the X-ray Telescope and Science Operations Teams for NASA's Neil Gehrels Swift Observatory. He is a leading researcher in transient astrophysics including discovery of outbursting Stellar Mass Black Holes in binary systems, Be/X-ray Binaries, Gamma-Ray Bursts and Multi-messenger Astrophysics, leading searches for electromagnetic searches for counterparts of astrophysical sources detected by Gravitational Wave and Neutrino detectors. 

David Burrows is Professor of Astronomy and Astrophysics at Penn State University, where he has been involved in X-ray detector development and analysis of X-ray data from supernova remnants and gamma-ray bursts.  He is a Co-Investigator on the Chandra Advanced CCD Imaging Spectrometer (ACIS) instrument and led the team that built the Swift X-ray Telescope. He was awarded the 2007 Bruno Rossi Prize and the 2009 Muhlmann Award as a member of the Swift team. 

\end{document}